\documentclass[letterpaper,twocolumn]{article}
\usepackage[utf8]{inputenc}
\usepackage[T1]{fontenc}
\usepackage[style=chem-acs]{biblatex}
\usepackage[compat=0.8]{yquant}
\usepackage[letterpaper,margin=1in]{geometry}
\usetikzlibrary{quotes,positioning,tikzmark,fit}
\usepackage{hyperref,amsmath,braket,subcaption,placeins,cuted,authblk,multirow,amssymb,booktabs,xr}
\usepackage{etoolbox}

\newcounter{eqncount}

\newcommand{\eqsRef}[1]{%
  \setcounter{eqncount}{0}
  \renewcommand*{\do}[1]{\stepcounter{eqncount}}
  \docsvlist{#1}
  Eq%
  \ifnum\value{eqncount}>1
    s
  \fi
  .\nobreakspace
  \def\eqnrefdelim{\def\eqnrefdelim{, }}%
  \renewcommand{\do}{\eqnrefdelim\ref}%
  \textup{(}%
    \docsvlist{#1}%
  \textup{)}%
}

\author[1,2]{Albert J. Pool}
\author[1,2]{Michael Schelling}
\author[1,2,3]{Birger Horstmann*}
\affil[1]{Institute of Engineering Thermodynamics, German Aerospace Center (DLR), Wilhelm-Runge-Str.\ 10, 89081 Ulm, Germany}
\affil[2]{Helmholtz Institute Ulm, Helmholtzstr.\ 11, 89081 Ulm, Germany}
\affil[3]{Department of Physics, Ulm University, Albert-Einstein-Allee 11, 89081 Ulm, Germany}

\title{Simulation of a Battery Cell on Quantum Computers: Reactions \& Transport}
\date{*Email: birger.horstmann@dlr.de}

\begin{document}
\newcommand{\C}{\ensuremath{\hat{C}}}
\newcommand{\A}{\ensuremath{\hat{A}}}
\newcommand{\U}{\ensuremath{\hat{U}}}

\maketitle
\begin{abstract}
Simulations of electrochemical materials and systems accelerate technological progress, but are still limited by computational power.
In particular, quantum computing offers prospects for higher resolutions, due to the exponential amount of data that can be stored in a quantum state.
As current quantum computers are still noisy, we consider a hybrid quantum-classical algorithm, that divides the problem into smaller computational tasks.
We describe how to implement such an algorithm for non-linear partial differential equations, the Feynman--Kitaev Hamiltonian, in a scalable way for an electrochemical system.
We show how it can evaluate general electrochemical models and present a quantum simulation of the Single Particle Model with electrolyte (SPMe) as the first quantum simulation of a battery cell.
\end{abstract}

\begin{figure}[h]
    \centering
    \includegraphics{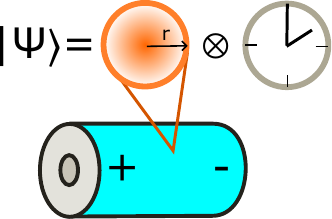}
    
    Graphical abstract
\end{figure}


Partial differential equations (PDEs) build the fundament of basically every discipline in science and engineering. 
In physical chemistry, PDEs are necessary to describe and simulate phenomena such as diffusion and reaction rates accurately, both for theoretical insight and for practical applications\autocite{sokalski2003numerical,kostre2021coupling,latz2011thermodynamic,braun2015thermodynamically,thomas2016classical,ruiz2024corrosion}.
The demand for more accurate computational results increases steadily, leading to higher and higher demands in both complexity and resolution. 
Especially multi-scale and three-dimensional models operate on the limits of classical methods\autocite{tadmor2012review}.
While Moore's Law has guaranteed the steady increase of computational power over the last decades, miniaturisation is reaching its physical limit\autocite{shalf2020future}, and the quest for novel ideas seems more relevant than ever.

In this context, quantum computing holds tremendous promises for solving PDEs\autocite{GiviEtAl20,tennie2025quantum}
by offering an exponentially large Hilbert state space of the physical qubits, whose potential might be harnessed to achieve high-resolution descriptions of physical chemistry models. 
Recently, a quantum algorithm for diffusion through a fuel cell membrane has been presented.\autocite{gubaev2026variational}
As quantum computers inherently perform unitary, and thus linear, operations, special consideration have to be made to achieve scalable algorithms for non-linear problems,
for which various methods have been recently proposed and tested\autocite{jin2024analog,lu2024quantum,berger2025trainable,costa2025further,tennie2025integration,siegl2026tensor}.

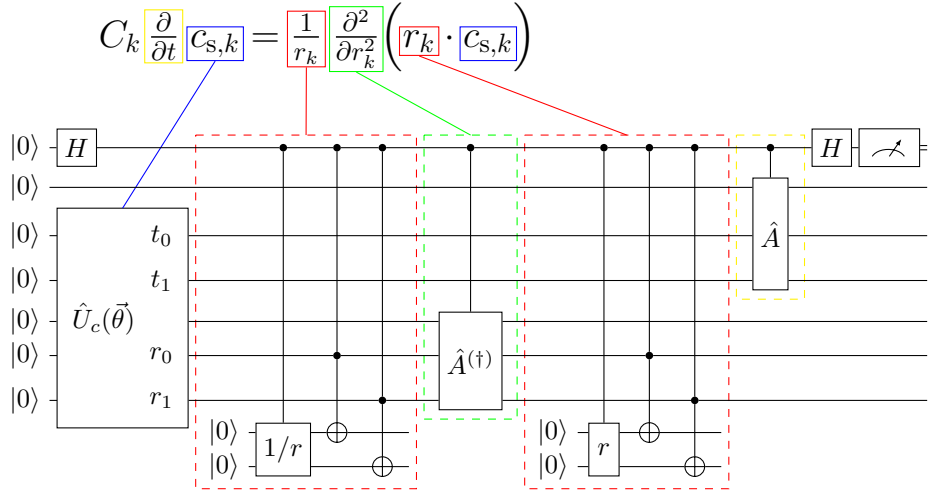
\begin{figure*}[t]
    \centering
    \begin{tikzpicture}[remember picture]
    \begin{yquant}
        \node (eq) at (3.54,0) {\Large $C_k\,
        \subnode[inner sep=1pt]{dt_sub}{\frac{\partial}{\partial t}}\;
        \subnode[inner sep=1pt]{c_sub}{c_{\text{s},k}}
        = 
        \subnode[inner sep=1pt]{r_sub}{\frac{1}{r_k}}\;
        \subnode[inner sep=1pt]{d_sub}{\frac{\partial^2}{\partial r _k^2}}
        \Big(
        \subnode[inner sep=1pt]{rinv_sub}{r_k}
        \cdot
        \subnode[inner sep=1pt]{c_sub2}{c_{\text{s},k}}
        \Big)$};
    \node (dt_node) [fit=(dt_sub),draw,style={yellow},inner sep=0pt] {};
    \node (c_node) [fit=(c_sub),draw,style={blue},inner sep=0pt] {};
    \node (c_node2) [fit=(c_sub2),draw,style={blue},inner sep=0pt] {};
    \node (r_node) [fit=(r_sub),draw,style={red},inner sep=0pt] {};
    \node (d_node) [fit=(d_sub),draw,style={green},inner sep=0pt] {};
    \node (rinv_node) [fit=(rinv_sub),draw,style={red},inner sep=0pt] {};
    nobit nob[3];
    qubit {$\ket{0}$} a;
    qubit {$\ket{0}$} at;
    qubit {$\ket{0}$} t[2];
    qubit {$\ket{0}$} ax;
    qubit {$\ket{0}$} x[2];
    h a[0];
    
    [name=U]
    subcircuit {
    \yquantset{operator/separation=0pt}
    qubit {} t[2];
    qubit {} x[2];
    discard -;
    text {$\U_c(\vec{\theta})$} (t,x);
    [inner xsep=0pt, outer xsep=0pt, minimum width=0pt]
    inspect {$t_\idx$} t;
    [inner xsep=0pt, outer xsep=0pt, minimum width=0pt]
    inspect {$r_\idx$} x;
    } (t,x);
    settype {qubit} x;
    settype {qubit} t;
    \draw (U.north) edge (c_node.south) [style={blue}];
    
    [this subcircuit box style={dashed,red},name=r]
    subcircuit {
    qubit {} a;
    qubit {} at;
    qubit {} ax;
    qubit {} t[2];
    qubit {} x[2];
    [ancilla]
    qubit {$\ket{0}$} r[2];
    box {$1/r$} (r) | a;
    cnot r[0] | a, x[0];
    cnot r[1] | a, x[1];
    } (a,at,ax,t,x);
    \draw (r.north) edge (r_node.south) [style={red}];
    
    [this subcircuit box style={dashed,green},name=d]
    subcircuit {
    qubit {} a;
    qubit {} at;
    qubit {} t[2];
    qubit {} ax;
    qubit {} x[2];
    box {$\A^{(\dagger)}$} (ax, x) | a;
    } (a,at,t,ax,x);
    \draw (d.north) edge (d_node.south) [style={green}];
    
    [this subcircuit box style={dashed,red},name=rinv]
    subcircuit {
    qubit {} a;
    qubit {} at;
    qubit {} ax;
    qubit {} t[2];
    qubit {} x[2];
    [ancilla]
    qubit {$\ket{0}$} rinv[2];
    box {$r$} (rinv) | a;
    cnot rinv[0] | a, x[0];
    cnot rinv[1] | a, x[1];
    } (a,at,ax,t,x);
    \draw (rinv.north) edge (rinv_node.south) [style={red}];
    
    [this subcircuit box style={dashed,yellow},name=dt]
    subcircuit {
    qubit {} a;
    qubit {} at;
    qubit {} t[2];
    box {$\A$} (at, t) | a;
    } (a,at,t);

    h a[0];
    measure a;
    
    \end{yquant}
    \end{tikzpicture}
    \caption{ 
        One of the quantum circuits encoding Eq.~\eqref{eq:derivative}, showcasing the quantum implementation of a differential equation for two qubits in space and time.
        The ansatz circuit $\U_c(\vec{\theta})$ (blue) encodes the concentration in time and space. The multiplication with $r$ and $1/r$ (red) realises the transformation to spherical coordinates. The binary shift operator $\A^{(\dagger)}$ (green) implements the spatial derivative; the dagger in parentheses indicates that both the left and right derivative are required (in separate circuits) to determine the Laplacian. The shift operator $\A$ on the right (yellow) implements the time derivative. Both have an ancilla qubit as their most significant bit to remove periodicity. This circuit implements the spherical Laplacian using the discretisation described in Sec.~1 of the supporting information.
        }
    \label{fig:circ-example}
\end{figure*}

This work presents the first solution for a full electrochemical cell using a quantum algorithm.
Our method employs a combination of the Feynman--Kitaev (FK) Hamiltonian\autocite{jarrod_mcclean_feynmans_2013,tempel2014kitaev,barison2022variational} with Quantum Nonlinear Processing Units (QNPUs)\autocite{lubasch2020variational,sarma2024quantum}, previously used for the Burgers equation\autocite{pool2024nonlinear}.
We extend this with an improved, scalable treatment of boundary conditions, as well as a novel treatment of coupled nonlinear equations.
The FK Hamiltonian gives a spacetime formulation, where the state of the system at all times and points in space results from a single optimisation process.
This formulation allows to extend the exponential scaling advantage to the time domain and circumvents the accumulation of errors of time-stepping methods where the state needs to be read out after every step.
Combining this spacetime FK formulation with a variational approach makes it compatible with current noisy intermediate-scale quantum (NISQ) hardware\autocite{cerezo2021variational,jaksch2023variational}, while the FK formulation itself can also be incorporated into quantum algorithms for future fault-tolerant quantum hardware.
We consider the implementation on today's superconducting hardware with an experiment on an emulated IBM superconducting computer.
We believe our work provides a solid foundation for more complex systems, opening the way for studying 3D models and more sophisticated differential-algebraic systems of equations using quantum computing.

The main component of a battery cell model is
a differential-algebraic system of equations that describes the transport of charge carriers (e.g.\ lithium ions) between the two electrodes of the cell as well as the chemical reactions taking place on the electrode surfaces. 
The electrodes are porous structures consisting of spherical particles. Here, the pores are filled with electrolyte, which allows ions to be transported through these pores and reach every electrode particle.
There are different established methods to simulate these electrode particles. In the Doyle--Fuller--Newman (DFN) model\autocite{doyle1993modeling}, the electrodes consist of a 1D grid where a radially symmetric particle is modelled at every grid point, while in the Single Particle Model with electrolyte (SPMe)\autocite{marquis2019asymptotic}, only one representative particle is modelled for each electrode.
In order to determine the electrochemical potential in the electrolyte as well as the electrode potentials, a galvanostatic condition (fixed charge/discharge current) or potentiostatic condition (fixed cell voltage) is used.
The electrode and electrolyte variables are coupled to each other by the Butler--Volmer equation, which gives the reaction rate on the electrode particle surface as a function of these variables.

\begin{table*}[t]
  \centering
  \caption{Physical parameters of the described battery.}
  \label{tab:phys-parameters}
  \begin{tabular}{cp{0.32\textwidth}ccc}
  Parameter & Description
  & anode & separator & cathode \\ \midrule
  $D_{\text{s},k}^*$ & Diffusion coefficient ($k\in\{\text{n},\text{p}\}$) & $2\times10^{-16}$ m\textsuperscript{2}/s~\textsuperscript{(d)} & & $5\times10^{-13}\text{~m}^2/\text{s}$~\textsuperscript{(a)} \\
  $R_k^*$ & Particle radius ($k\in\{\text{n},\text{p}\}$) & 80 nm~\textsuperscript{(c)} & & 3.8 µm~\textsuperscript{(a)} \\
  $C_A^*$ & Areal capacity & 6.5 Ah/m\textsuperscript{2}~\textsuperscript{(d)} & & 6.5 Ah/m\textsuperscript{2}~\textsuperscript{(b)} \\
  $L_k^*$ & Thickness ($k\in\{\text{n},\text{s},\text{p}\}$) & 4~µm~\textsuperscript{(d)} & 12~µm~\textsuperscript{(a)} & 15~µm~\textsuperscript{(b)} \\
  $\epsilon_k$ & Porosity ($k\in\{\text{n},\text{s},\text{p}\}$) & 0.561~\textsuperscript{(c)} & 0.45~\textsuperscript{(a)} & 0.345~\textsuperscript{(b)} \\
  $\beta$ & Bruggeman coefficient & 1.5 & 1.5 & 1.5 \\
  $D_\text{e}^*$ & Electrolyte diffusion constant & \multicolumn{3}{c}{$4.6\times10^{-10}\text{~m}^2/\text{s}$~\textsuperscript{(e)}} \\
  $\kappa^*$ & Electrolyte ionic conductivity & \multicolumn{3}{c}{1.16 S/m\textsuperscript{2}~\textsuperscript{(e)}} \\
  $c_\text{e}^0$ & Electrolyte standard concentration & \multicolumn{3}{c}{1 M} \\
  $t_+$ & Transference number & \multicolumn{3}{c}{0.38~\textsuperscript{(e)}} \\
  \midrule
  \multicolumn{5}{l}{
    \textsuperscript{(a)}~Cathode material and separator parameters from Sturm et al.\autocite{STURM2019204}
  } \\
  \multicolumn{5}{l}{\multirow{2}{.95\textwidth}{\raggedright
    \textsuperscript{(b)}~Cathode geometry parameters from Wu et al.\autocite{wu2026integrative} with thickness and capacity divided by 4 for a high-power cell
  }} \\
  \multicolumn{5}{l}{}\\
  \multicolumn{5}{l}{
    \textsuperscript{(c)}~Anode parameters from Pan et al.\autocite{pan2019systematic}
  } \\
  \multicolumn{5}{l}{\multirow{2}{.95\textwidth}{\raggedright
    \textsuperscript{(d)}~Estimated for a matching diffusion timescale and areal capacity of both electrodes. $D_{\text{s},\text{n}}$ is within the range given by Pan et al.\autocite{pan2019systematic}
  }} \\
  \multicolumn{5}{l}{}\\
  \multicolumn{5}{l}{
    \textsuperscript{(e)}~Electrolyte parameters from molecular dynamics simulations of Lehnert et al.\autocite{lehnert2025combining}
  }
  \end{tabular}
\end{table*}

\begin{table*}[t]
  \centering
  \caption{Simulation parameters for the nondimensionalised SPMe.}
  \label{tab:sim-parameters}
  \begin{tabular}{cp{0.49\textwidth}cc}
  Parameter & Description
  & anode & cathode \\ \midrule 
  $c_0$ & Initial concentration of the electrodes, as fraction of the maximum concentration & 0.1 & 0.9 \\ 
  $j_k$ & \hspace{0pt}Dimensionless current density ($k\in\{\text{n},\text{p}\}$) & 0.0125 & -0.0125\\ 
  & & \multicolumn{2}{c}{$\triangleq4.13$C} \\
  $k_{\text{f},k}$ & Dimensionless reaction rate parameter ($k\in\{\text{n},\text{p}\}$) & \multicolumn{2}{c}{0.125} \\
  $D_{\text{s},k}$ & Dimensionless diffusion coefficient ($k\in\{\text{n},\text{p}\}$) & \multicolumn{2}{c}{1} \\
  \bottomrule
  \end{tabular}
\end{table*}

As demonstration case for our quantum algorithm, we solve the SPMe model with galvanostatic conditions, which consists of a set of differential equations for concentration inside the electrode particles and a set of analytical expressions for concentration and electrochemical potential in the electrolyte.
We formulate these equations in such a way that they can be efficiently implemented in quantum circuits,
as detailed in Sec.~1 of the supporting information\autocite{zeng2013efficient,iten2016quantum,siegl2026tensor}.
We describe the concentration inside each electrode particle by the diffusion equation in spherical coordinates, Eq.~(40a) from Marquis et al.\autocite{marquis2019asymptotic}, which we rewrite as
\begin{equation}
    \frac{\partial c_{\text{s},k}}{\partial t} = {D_{\text{s},k}}
        \frac{1}{r_k}\frac{\partial^2}{\partial r _k^2}\Big(r_k\cdot c_{\text{s},k}\Big),\label{eq:derivative}
\end{equation}
where $k\in\{\text{n},\text{p}\}$ stands for the (negative) anode or (positive) cathode, 
$c_{\text{s},k}=c^*_{\text{s},k}/c^\text{max}_{\text{s},k}$ is the dimensionless solid concentration inside the electrode particle, obtained by dividing the real concentration through the maximum of the respective electrode material, $r_k=r^*_k/R_k^*$ the radial coordinate nondimensionalised by the particle radius $R_k^*$, and $D_{\text{s},k}=D^*_{\text{s},k}\tau_\text{d}/R_k^{*2}$ the nondimensionalised diffusion coefficient. Here, $\tau_\text{d}$ is the characteristic timescale of diffusion, which we also use as the unit of time in the simulation, and choose it to be the same for both electrodes ($\tau_{\text{d},\text{n}}=\tau_{\text{d},\text{p}}=28.8$~s to obtain $D_{\text{s},k}=1$). The boundary conditions are given by Eq.~(40d) from Marquis et al.\autocite{marquis2019asymptotic} as
\begin{align}
    \frac{\partial c_{\text{s},k}}{\partial r_{\text{s},k}}\Big|_{r_k=0} = 0,
    & &   -D_{\text{s},k}\frac{\partial c_{\text{s},k}}{\partial r_{\text{s},k}}\Big|_{r_k=1} = j_k,\label{eq:boundary}
\end{align}
where $j_k$ is a nondimensionalised fixed current (galvanostatic condition).

\begin{figure*}[t]
    \begin{subfigure}[b]{\textwidth}
        \begin{minipage}[c]{.16\textwidth}
        \captionsetup{justification=raggedleft,singlelinecheck=off}
        \caption{}
        \label{fig:charge-battery}
        \end{minipage}
        \begin{minipage}[c]{.75\textwidth}
        \includegraphics[scale=.75]{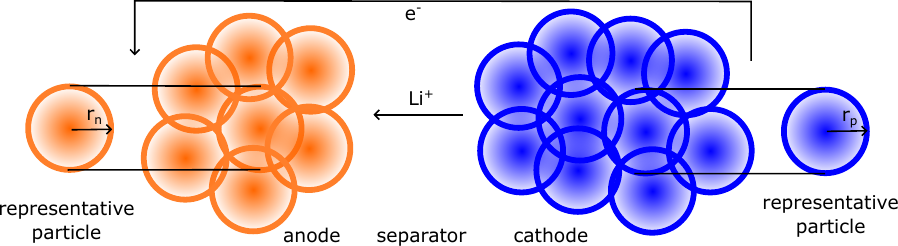}
        \end{minipage}
    \end{subfigure}
    \begin{subfigure}{\textwidth}
        \includegraphics[trim=4mm 4mm 3mm 3mm,clip]{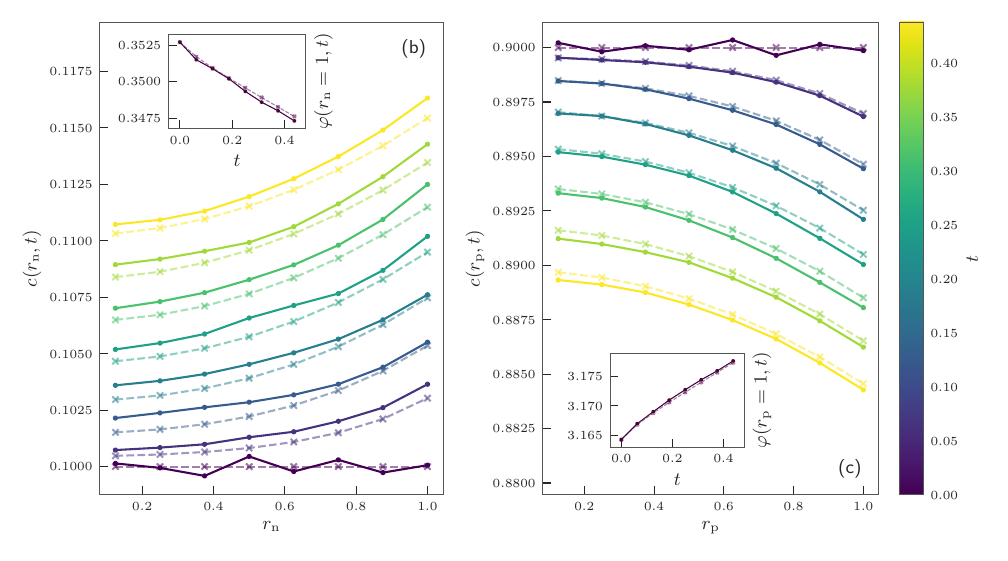}
        \phantomcaption\label{fig:charge-anode}
        \phantomcaption\label{fig:charge-cathode}
    \end{subfigure}
    \caption{Concentration and potential for the anode and cathode for a charging process with a constant current density $j_\text{n}=0.0125$/$j_\text{p}=-0.0125$ and $\Delta t=1/16$, starting at an anode concentration of $0.1$ and a cathode concentration of $0.9$.
    (a) Schematic explaining the two electrodes and representative particles, inspired by\autocite{kuhn2023bayesian}.
    (b) Concentration and electrochemical potential $\varphi_\text{n}=\phi_{\text{s},\text{n}}-\phi_{\text{e},\text{n}}$ (inset) in the anode representative particle as a function of time and space, with an infidelity of $2.8\times10^{-6}$ for the concentration.
    (c) Concentration and electrochemical potential $\varphi_\text{p}=\phi_{\text{s},\text{p}}-\phi_{\text{e},\text{p}}$ (inset) in the cathode representative particle as a function of time and space, with an infidelity of $2.1\times10^{-5}$ for the concentration. In both figures, the dashed lines represent a numerical solution to the model equations.}
    \label{fig:charge-result}
\end{figure*}

For a scalable quantum implementation, we modify the outer boundary condition such that the reaction is applied in a small boundary region $r\geq r_\text{r}$ instead of a boundary point, as detailed in Sec.~1.1 of the supporting information.
As a proof of concept, we take the Butler--Volmer equation, Eq.~(3g/i) from Marquis et al.\autocite{marquis2019asymptotic}, and linearise it as
\begin{subequations}
\begin{align}
    R(c_k,\varphi_k) &= k_{\text{f},k} (
        a_0+a_1\cdot c_k + a_2\cdot \varphi_k \nonumber \\
        & + a_3\cdot c_k \cdot \varphi_k )/V_\text{th}, \label{eq:bv-approx}\\
        &\approx j_{0,k} \sinh \left(\frac{\varphi-U_k\left(c_{\text{s},k}\right)}{2V_\text{th}}\right),\label{eq:bv}
\end{align}
\end{subequations}
with $\varphi_k=\phi_{\text{s},k}-\phi_{\text{e},k}$ and $c_k=c_{\text{s},k}$, where $\phi_{\text{s},k}$ and $\phi_{\text{e},k}$ are the potential of each electrode $k\in\{\text{n},\text{p}\}$, and of the electrolyte in its pores, respectively, $U_k$ is the open circuit voltage (OCV) function of the electrode material under consideration, and $V_\text{th}=RT/F$ the thermal voltage.
$k_{\text{f},k}$ is the reaction rate constant, a prefactor extracted from the concentration-dependent exchange current density $j_{0,k}$ for each electrode.
We implement the galvanostatic condition in an algebraic equation
\begin{equation}
    \sum_{r\geq r_\text{r}} R(c_k(r),\varphi_k(r)) \Delta r = j_k \quad \forall t,\label{eq:AE}
\end{equation}
which couples $\varphi_k$ and $c_k$. 
We note that the SPMe model allows calculating the potentials $\varphi_k$ in postprocessing; accordingly, it would be possible to only model the concentration with quantum circuits and determine all potentials analytically.
However, we deliberately choose to implement the coupled equations in our quantum algorithm to showcase this important step towards more sophisticated models.

\begin{figure*}[t]
    \centering
    \begin{subfigure}{\textwidth}
        \includegraphics[trim=4mm 4mm 3mm 3mm,clip]{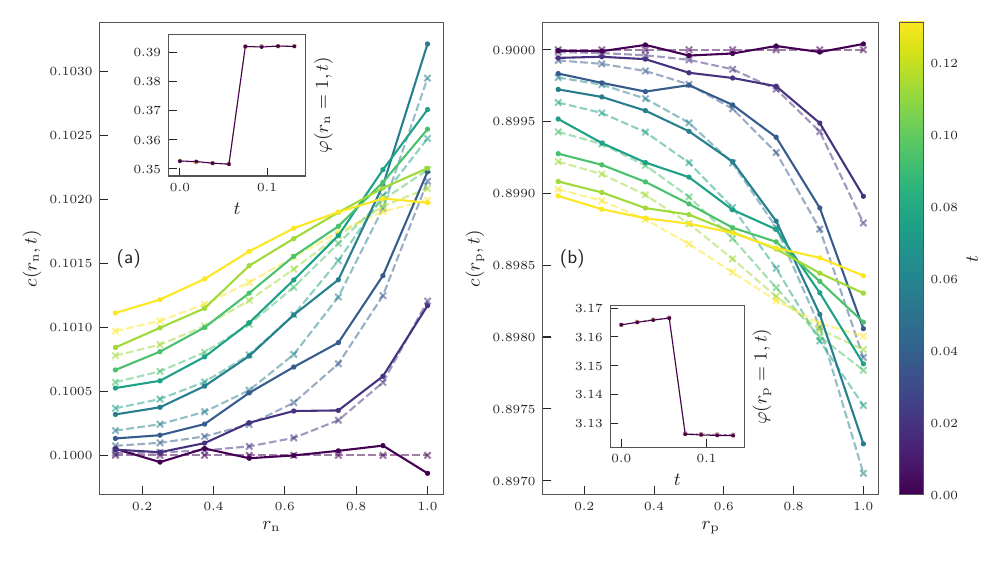}
        \phantomcaption\label{fig:gitt-anode}
        \phantomcaption\label{fig:gitt-cathode}
    \end{subfigure}
    \caption{Concentration and potential for the anode and cathode for a GITT pulse with a duration of $0.075$ and an equally long relaxation phase, starting at an anode concentration of $0.1$ and a cathode concentration of $0.9$. (a) Anode concentration and electrochemical potential $\varphi_\text{n}=\phi_{\text{s},\text{n}}-\phi_{\text{e},\text{n}}$ (inset) as a function of time and space, with an infidelity of $4.5\times10^{-7}$ for the concentration. (b) Cathode concentration and electrochemical potential $\varphi_\text{p}=\phi_{\text{s},\text{p}}-\phi_{\text{e},\text{p}}$ (inset) as a function of time and space, with an infidelity of $3.0\times10^{-7}$ for the concentration. In all figures, the dashed lines represent a numerical solution to the model equations.}
    \label{fig:gitt-result}
\end{figure*}

We attempt to solve the system of~\eqsRef{eq:derivative,eq:boundary,eq:AE} of the SPMe
with a quantum algorithm.
For an exponential scaling in both time and space we choose the FK Hamiltonian\autocite{tempel2014kitaev}, which encodes the full space-time solution in its ground state.
To make our approach feasible for NISQ devices, we use a hybrid quantum-classical optimisation, where the quantum computer evaluates a quadratic cost function and the optimisation is performed on classical hardware.
We extend our existing implementation\autocite{pool2024nonlinear} with quantum circuits for spherical diffusion, Neumann boundary conditions, as well as reaction terms with a coupling between the concentration and potential.
For the spherical diffusion, we use quantum nonlinear processing units (QNPUs)\autocite{lubasch2020variational} to multiply with quantum circuits that represent the functions $r_k$ and $1/r_k$ from Eq.~\eqref{eq:derivative}. An example circuit is shown in Fig.~\ref{fig:circ-example}. We make this system non-periodic by adding one more spatial qubit, and introduce an efficient implementation of the Neumann boundary condition, that significantly reduces the amount of circuits and multi-qubit gates compared to the existing implementation by Over et al.\autocite{over2025boundary}, which is detailed in Sec.~2.1 of the supporting information.
Lastly, we add QNPU circuits to calculate all terms of Eq.~\eqref{eq:bv-approx}. We then use similar circuits, without any time evolution, to implement the algebraic equation~\eqref{eq:AE}.
All of these are detailed in Sec.~2 of the supporting information.

After obtaining a solution with our quantum algorithm for an SPMe of both an anode and a cathode particle, we can extract the cell voltage of the simulated battery cell.
We determine the overpotentials $\Bar{\eta}_c$ and $\Delta\Bar{\phi}_\text{e}$ due to electrolyte transport as described in Sec.~3 of the supporting information\autocite{marquis2019asymptotic}.
We then calculate the cell voltage from the electrochemical potentials $\varphi_k$ of both electrodes as
\begin{equation}
    U=\varphi_\text{p}-\varphi_\text{n}+\Bar{\eta}_c+\Delta\Bar{\phi}_\text{e}.
\end{equation}
We nondimensionalise the required electrolyte parameters in a similar fashion to the electrode parameters. We divide the liquid concentration $c_{\text{e},k}=c^*_{\text{e},k}/c^0_\text{e}$ by its standard value $c^0_\text{e}$ and the length scale by the thickness of the cell, such that $L_\text{p}+L_\text{s}+L_\text{n}=1$.

For a proof of principle of the quantum algorithm, we assume that both electrodes have the same areal capacity and timescale of diffusion $\tau_\text{d}$.
We implement linearised electrode characteristics as described in Sec.~4 of the supporting information, taking sample parameters $a_0$ to $a_3$ in Eq.~\eqref{eq:bv-approx} from a silicon anode\autocite{pan2019systematic} and NMC811 cathode\autocite{STURM2019204}, respectively. We model a high-power cell with thin electrodes and a current density just above 4C.
We assume a reaction rate parameter of $k_{\text{f},k}=0.125$ for both electrodes, for a sufficiently large coupling between potential and concentration, in order to observe the effect of an increasing/decreasing charging state in the cell voltage.
This corresponds to a cell with the parameters given in Tab.~\ref{tab:phys-parameters}.
We then use a noiseless simulator of a quantum computer to optimise a parameterised quantum circuit, in order to find the minimum of the FK Hamiltonian, which represents the solutions for $c_k$ and $\varphi_k$.
To facilitate this variational optimisation, we employ an adaptive optimisation strategy using the Adam\autocite{kingma2014adam} and L-BFGS-B\autocite{byrd1995limited} optimisers and a multigrid approach to refine the spatial grid\autocite{lubasch2018multigrid}, as detailed in Sec.~5 of the supporting information.

We validate our method by simulating two different operation modes.
In the charging case, we apply a galvanostatic condition and simulate long enough that a constant concentration gradient inside the particle is reached.
In the GITT (galvano\-static inter\-mittent titration technique)\autocite{weppner1977determination} case, we simulate a short pulse with constant current followed by an equally long relaxation phase with zero current. This case highlights the dynamics at short timescales, where a concentration gradient is formed and disappears again, which is recognisable in the cell voltage due to the coupling between concentration and potential.
The parameters for both of these cases are given in Tab.~\ref{tab:sim-parameters}. We perform all simulations with $3+3$ qubits, meaning $2^3=8$ points in both time and space.

\begin{figure}[t]
    \centering
    \includegraphics[trim=4mm 4mm 3mm 3mm,clip]{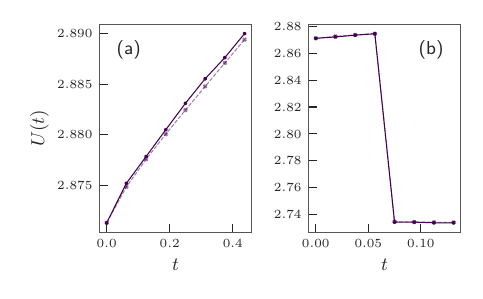}
    \caption{Cell voltage for (a) charging with a constant current and $\Delta t=1/16$ and (b) a GITT pulse lasting $0.075$~s followed by an equally long relaxation phase. The dashed lines represent a numerical solution to the model equations.}
    \label{fig:cell-voltage}
\end{figure}

First, we simulate the charging case. The resulting concentrations and potentials are shown in  Fig.~\ref{fig:charge-result}.
The modelled anode particle clearly shows the characteristics of a charging scenario, with a lithiation of the particle, i.e.\ increasing concentration at its surface $r_\text{p}=1$ and diffusion towards the particle centre, and a drop in the potential over time.
The modelled cathode particle exhibits an inverse behaviour, showing the characteristics of delithiation and an associated surge of the electrode potential.
We observe a parabolic shape of the lithium concentration, which validates the quantum implementation of the spherical diffusion, as the concentration would evolve towards a straight line in Cartesian coordinates.
Both the closed Neumann boundary condition in the particle centres, $r_k =0$, and the reactions on the outer boundaries $r_k =1$ are captured by our circuits.
While the general shape is adequately met, the reaction overshoots slightly and the initial condition is not fully matched. 
The potential is matched very well for both electrodes, showcasing the successful coupling of the two variables through Eq.~\eqref{eq:bv-approx} by our method.

In the second experiment, we consider the GITT case, for which the results are shown in Fig.~\ref{fig:gitt-result}.
Again, the obtained solution shows the expected characteristics, displaying the lithiation of the anode and the delithiation of the cathode particle.
As in the previous experiment both the spherical geometry, here in the parabolic shape of the intermediate time steps, and the closed boundary in the particle centre are visible.
On the outer boundary, $r_k = 1$, the influence of the sudden cut of the external voltage is clearly visible by a decrease in concentration during the later time steps, which extends into the particle over time.
This relaxation of the concentration profiles indicate that the corresponding drop in the reaction rates at the particle surfaces is captured. 
This is assured by the sudden change in the particle potentials, which are again met very well.
We see that the concentration in the second half of time shows some deviation from the expected profile, being too high in the centre of the particle for the anode case and too low at the edge in the cathode case.
We note that a relaxation of the potential is not visible at this timescale.

We derive the resulting cell voltage for the two cases, charging and GITT, as can be seen in Fig.~\ref{fig:cell-voltage}.
The voltage depicts the difference between the two potentials plus an electrolyte contribution that depends on the current and properties of the cell.
As can be seen in both figures, the voltage is well reproduced, showing the same trends as the potentials of the individual electrodes, demonstrating that the cell voltage can be successfully retrieved from the proposed quantum algorithm in both scenarios.

\begin{figure*}[t]
    \centering
    \begin{subfigure}{\textwidth}
        \includegraphics[trim=4mm 4mm 0 3mm,clip]{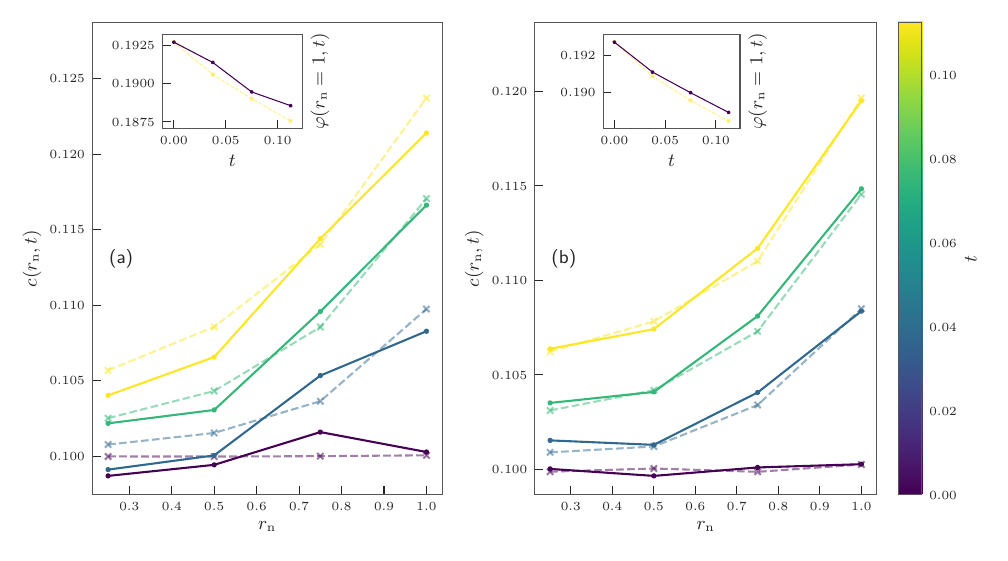}
        \phantomcaption\label{fig:ibm_fake_berlin_full}
        \phantomcaption\label{fig:ibm_fake_berlin_reduced}
    \end{subfigure}
    \caption{Comparison of two different circuits with $2+2$ qubits ($2^2=4$ points in space and time) on the IBM Berlin fake backend, showing concentration as a function of time and space and (inset) potential as a function of time. (a) 4 layers with all-to-all connectivity. (b) 3 layers hardware-efficient, only including connections that are physically present. The dashed lines correspond to a result from a noiseless simulator for the same circuits. Both results were produced using 2.5 million shots with M3 error mitigation.}
    \label{fig:ibm_fake_berlin}
\end{figure*}

Lastly, we prepare a circuit with less qubits and gates suitable for current superconducting quantum computing hardware. We take an optimised circuit for $2+2$ qubits for the anode in the charging case, with a higher charging current of $j_\text{n}=0.0625$ to have larger differences in amplitude, making it less susceptible to noise. We start with an optimised circuit where all four qubits are connected to each other in four layers, and progressively reduce this to three layers where only neighbouring qubits are connected by removing connections while optimising again. We sample both circuits on the IBM Berlin fake backend, which emulates an IBM Quantum Nighthawk processor with a square topology.
We perform 2.5 million shots with M3 error mitigation\autocite{nation_scalable_2021} and show the results in Fig.~\ref{fig:ibm_fake_berlin}.
Comparing the left and right side of the figure, it is apparent that the circuit with a lower gate count produces less noise.
We also show a similar experiment on the IQM Emerald\autocite{abdurakhimov2024technology} superconducting quantum computer in Sec.~6 of the supporting information.

In summary, we demonstrate the solution of a full electrochemical model, consisting of a system of coupled nonlinear equations, using a quantum algorithm. 
For this purpose we transformed the electrochemical processes of the SPMe into a quantum algorithm, representing boundary conditions, coupled variables and nonlinear terms with quantum circuits. 
The experiments demonstrate that our chosen approach is functional, captures the characteristics of the simulated battery electrodes, that relevant battery parameters such as the cell voltage can be extracted, and that it can be extended to include coupled equations.
Using a variational approach based on the FK Hamiltonian, our method is scalable as the depth of the quantum circuits scales linearly with the number of qubits and the number of circuits only depends on the equations and desired precision. 
While we employ strategies to speed up the variational optimisation by using a multigrid approach\autocite{pool2024nonlinear}, the optimisation process currently remains a practical bottleneck of unclear complexity.
However, it is possible to improve scalability by using other methods to find the ground state such as subspace solvers\autocite{kirby2023exact} or quantum imaginary time evolution\autocite{mao2023measurement}.
Due to the limitations of current quantum hardware, our experiments remain small-scale demonstrations.
Additional experiments may be conducted with a higher spatial resolution, in order to resolve larger concentration gradients inside the particles for a better simulation of the relaxation phase in the GITT case.
On the emulator of the IBM superconducting platform, we observe that the ansatz circuit needs to be reduced to be more hardware-efficient in order to sample the state, but this counteracts trainability. While superconducting quantum computers offer very fast gate times, they do have more limited connectivity than other platforms such as neutral atoms or trapped ions\autocite{tennie2025quantum}. This is not only a limit to variational algorithms, but also to error correcting codes for fault-tolerant algorithms\autocite{campbell2017roads}.
Some recent developments aim to alleviate this, such as multi-qubit couplers\autocite{feulner2026quantum} or architectures with a central resonator\autocite{renger2026superconducting}.

Conclusively, in providing a quantum simulation of an electrochemical cell using a Hamiltonian formulation, we believe our work can serve as a basis for more advanced models, and we see no fundamental limitation to extend it to more complex electrochemical models, such as the two-dimensional DFN model\autocite{doyle1993modeling} or microstructure-resolved 3D models\autocite{hein2016stochastic}, or other systems of coupled non-linear PDEs in chemistry or engineering.

\section*{Methods}
We implement all quantum circuits in Qiskit\autocite{javadi2024quantum} and verify that they are numerically equivalent to the functions we want to evaluate. To speed up the optimisation, we generate a statevector for the ansatz circuit in every optimisation step and calculate the cost function classically. 
All results represent the best out of 20 random initial conditions for every optimisation procedure.
We emulate the IBM Berlin superconducting quantum computer using the fake backend in the Qiskit Runtime library.

\section*{Acknowledgements}
We thank Alexios A. Michailidis for helpful discussions about implementation details and Lukas Köbbing and Johannes Hörmann for providing the OCV functions of the electrode materials.

This work is part of the project `Battery materials simulation using quantum computers' (BASIQ), \href{https://qci.dlr.de/en/basiq/}{https://qci.dlr.de/en/basiq/}, which is made possible by the DLR Quantum Computing Initiative (QCI) and the German Federal Ministry of Research, Technology and Space (BMFTR). The classical computations were performed on the JUSTUS 2 cluster, supported by the state of Baden-Württemberg through bwHPC and the German Research Foundation (DFG) through Grant No. INST 40/575-1 FUGG.
This work contributes to the research performed at CELEST (Center for Electrochemical Energy Storage Ulm-Karlsruhe).

\switchonecolumn
\FloatBarrier
\printbibliography

@article{marquis2019asymptotic,
  title={An asymptotic derivation of a single particle model with electrolyte},
  author={Marquis, Scott G and Sulzer, Valentin and Timms, Robert and Please, Colin P and Chapman, S Jon},
  journal={J. Electrochem. Soc.},
  volume={166},
  number={15},
  pages={A3693},
  year={2019},
  publisher={IOP Publishing}
}

@article{lubasch2020variational,
 title = {Variational quantum algorithms for nonlinear problems},
 author = {Lubasch, Michael and Joo, Jaewoo and Moinier, Pierre and Kiffner, Martin and Jaksch, Dieter},
 journal = {Phys. Rev. A},
 volume = {101},
 number = {1},
 pages = {010301(R)},
 year = {2020},
 publisher = {APS},
 doi = {10.1103/PhysRevA.101.010301}
}

@article{doyle1993modeling,
  title={Modeling of galvanostatic charge and discharge of the lithium/polymer/insertion cell},
  author={Doyle, Marc and Fuller, Thomas F and Newman, John},
  journal={J. Electrochem. Soc.},
  volume={140},
  number={6},
  pages={1526},
  year={1993},
  publisher={IOP Publishing}
}

@article{pool2024nonlinear,
  title={Nonlinear dynamics as a ground-state solution on quantum computers},
  author={Pool, Albert J and Somoza, Alejandro D and Mc Keever, Conor and Lubasch, Michael and Horstmann, Birger},
  journal={Phys. Rev. Res.},
  volume={6},
  number={3},
  pages={033257},
  year={2024},
  publisher={APS}
}

@article{barison2022variational,
  title = {Variational dynamics as a ground-state problem on a quantum computer},
  author = {Barison, Stefano and Vicentini, Filippo and Cirac, Ignacio and Carleo, Giuseppe},
  journal = {Phys. Rev. Res.},
  volume = {4},
  number = {4},
  pages = {043161},
  numpages = {12},
  year = {2022},
  month = dec,
  publisher = {American Physical Society},
  doi = {10.1103/PhysRevResearch.4.043161},
}

@article{over2025boundary,
  title={Boundary treatment for variational quantum simulations of partial differential equations on quantum computers},
  author={Over, Paul and Bengoechea, Sergio and Rung, Thomas and Clerici, Francesco and Scandurra, Leonardo and de Villiers, Eugene and Jaksch, Dieter},
  journal={Comput. Fluids},
  volume={288},
  pages={106508},
  year={2025},
  publisher={Elsevier}
}

@article{lubasch2018multigrid,
 title = {Multigrid renormalization},
 author = {Lubasch, Michael and Moinier, Pierre and Jaksch, Dieter},
 journal = {J. Comput. Phys.},
 volume = {372},
 pages = {587--602},
 year = {2018},
 publisher = {Elsevier},
 doi = {10.1016/j.jcp.2018.06.065}
}

@article{tadmor2012review,
  title={A review of numerical methods for nonlinear partial differential equations},
  author={Tadmor, Eitan},
  journal={Bull. Amer. Math. Soc.},
  volume={49},
  number={4},
  pages={507--554},
  year={2012},
  doi={10.1090/S0273-0979-2012-01379-4},
}

@article{cerezo2021variational,
  title={Variational quantum algorithms},
  author={Cerezo, Marco and Arrasmith, Andrew and Babbush, Ryan and Benjamin, Simon C and Endo, Suguru and Fujii, Keisuke and McClean, Jarrod R and Mitarai, Kosuke and Yuan, Xiao and Cincio, Lukasz and others},
  journal={Nat. Rev. Phys.},
  volume={3},
  number={9},
  pages={625--644},
  year={2021},
  publisher={Nature Publishing Group UK London},
  doi={10.1038/s42254-021-00348-9},
}

@article{jarrod_mcclean_feynmans_2013,
	title = {Feynman’s clock, a new variational principle, and parallel-in-time quantum dynamics},
	journal = {Proc. Natl. Acad. Sci. U.S.A.},
    volume = {110},
	url = {https://www.pnas.org/doi/10.1073/pnas.1308069110},
	number = {41},
	urldate = {2023-02-22},
	author = {{Jarrod McClean} and {John A. Parkhill} and {Alán Aspuru-Guzik}},
	month = sep,
	year = {2013},
	doi = {10.1073/pnas.1308069110},
	pages = {E3901},
}

@article{GiviEtAl20,
author = {Givi, Peyman and Daley, Andrew J. and Mavriplis, Dimitri and Malik, Mujeeb},
title = {Quantum speedup for aeroscience and engineering},
journal = {AIAA J.},
volume = {58},
number = {8},
pages = {3715-3727},
year = {2020},
doi = {10.2514/1.J059183},
URL = {https://doi.org/10.2514/1.J059183}
}

@article{jaksch2023variational,
  title={Variational quantum algorithms for computational fluid dynamics},
  author={Jaksch, Dieter and Givi, Peyman and Daley, Andrew J and Rung, Thomas},
  journal={AIAA J.},
  volume={61},
  number={5},
  pages={1885--1894},
  year={2023},
  publisher={American Institute of Aeronautics and Astronautics},
  doi={10.2514/1.J062426},
}

@article{tennie2025quantum,
  title={Quantum computing for nonlinear differential equations and turbulence},
  author={Tennie, Felix and Laizet, Sylvain and Lloyd, Seth and Magri, Luca},
  journal={Nat. Rev. Phys.},
  volume={7},
  number={4},
  pages={220--230},
  year={2025},
  publisher={Nature Publishing Group UK London}
}

@article{shalf2020future,
  title={The future of computing beyond {Moore’s} {Law}},
  author={Shalf, John},
  journal={Philos. Trans. R. Soc. A},
  volume={378},
  number={2166},
  year={2020},
  publisher={The Royal Society}
}

@article{zeng2013efficient,
  title={Efficient conservative numerical schemes for {1D} nonlinear spherical diffusion equations with applications in battery modeling},
  author={Zeng, Yi and Albertus, Paul and Klein, Reinhardt and Chaturvedi, Nalin and Kojic, Aleksandar and Bazant, Martin Z and Christensen, Jake},
  journal={J. Electrochem. Soc.},
  volume={160},
  number={9},
  pages={A1565--A1571},
  year={2013},
  publisher={The Electrochemical Society}
}

@article{iten2016quantum,
  title={Quantum circuits for isometries},
  author={Iten, Raban and Colbeck, Roger and Kukuljan, Ivan and Home, Jonathan and Christandl, Matthias},
  journal={Phys. Rev. A},
  volume={93},
  number={3},
  pages={032318},
  year={2016},
  publisher={APS}
}

@article{tempel2014kitaev,
  title={The {Kitaev--Feynman} clock for open quantum systems},
  author={Tempel, David G and Aspuru-Guzik, Al{\'a}n},
  journal={New J. Phys.},
  volume={16},
  number={11},
  pages={113066},
  year={2014},
  publisher={IOP Publishing},
  doi={10.1088/1367-2630/16/11/113066},
}

@article{siegl2026tensor,
  title={Tensor-programmable quantum circuits for solving differential equations},
  author={Siegl, Pia and Reese, Greta Sophie and Hashizume, Tomohiro and van H{\"u}lst, Nis-Luca and Jaksch, Dieter},
  journal={Phys. Rev. Res.},
  volume={8},
  number={1},
  pages={013052},
  year={2026},
  publisher={APS}
}

@article{pan2019systematic,
  title={Systematic electrochemical characterizations of {Si} and {SiO} anodes for high-capacity {Li-Ion} batteries},
  author={Pan, Ke and Zou, Feng and Canova, Marcello and Zhu, Yu and Kim, Jung-Hyun},
  journal={J. Power Sources},
  volume={413},
  pages={20--28},
  year={2019},
  publisher={Elsevier}
}

@article{STURM2019204,
title = {Modeling and simulation of inhomogeneities in a 18650 nickel-rich, silicon-graphite lithium-ion cell during fast charging},
journal = {J. Power Sources},
volume = {412},
pages = {204-223},
year = {2019},
issn = {0378-7753},
doi = {https://doi.org/10.1016/j.jpowsour.2018.11.043},
url = {https://www.sciencedirect.com/science/article/pii/S0378775318312849},
author = {J. Sturm and A. Rheinfeld and I. Zilberman and F.B. Spingler and S. Kosch and F. Frie and A. Jossen}
}

@article{weppner1977determination,
  title={Determination of the kinetic parameters of mixed-conducting electrodes and application to the system {Li3Sb}},
  author={Weppner, W. and Huggins, Robert A},
  journal={J. Electrochem. Soc.},
  volume={124},
  number={10},
  pages={1569},
  year={1977},
  publisher={IOP Publishing}
}

@article{lu2024quantum,
  title={Quantum computing of reacting flows via {Hamiltonian} simulation},
  author={Lu, Zhen and Yang, Yue},
  journal={Proc. Combust. Inst.},
  volume={40},
  number={1-4},
  pages={105440},
  year={2024},
  publisher={Elsevier}
}

@article{kostre2021coupling,
  title={Coupling particle-based reaction-diffusion simulations with reservoirs mediated by reaction-diffusion {PDEs}},
  author={Kostr{\'e}, Margarita and Sch{\"u}tte, Christof and No{\'e}, Frank and Del Razo, Mauricio J},
  journal={Multiscale Model. Simul.},
  volume={19},
  number={4},
  pages={1659--1683},
  year={2021},
  publisher={SIAM}
}

@article{latz2011thermodynamic,
  title={Thermodynamic consistent transport theory of {Li-ion} batteries},
  author={Latz, Arnulf and Zausch, Jochen},
  journal={J. Power Sources},
  volume={196},
  number={6},
  pages={3296--3302},
  year={2011},
  publisher={Elsevier}
}

@article{sokalski2003numerical,
  title={Numerical solution of the coupled {Nernst}--{Planck} and {Poisson} equations for liquid junction and ion selective membrane potentials},
  author={Sokalski, Tomasz and Lingenfelter, Peter and Lewenstam, Andrzej},
  journal={J. Phys. Chem. B},
  volume={107},
  number={11},
  pages={2443--2452},
  year={2003},
  publisher={ACS Publications}
}

@article{ruiz2024corrosion,
  title={Corrosion modeling of aluminum alloys: a brief review},
  author={Ruiz-Garcia, A and Esquivel-Pe{\~n}a, V and God{\'\i}nez, FA and Montoya, R},
  journal={ChemElectroChem},
  volume={11},
  number={9},
  pages={e202300712},
  year={2024},
  publisher={Wiley Online Library}
}

@article{mao2023measurement,
  title={Measurement-based deterministic imaginary time evolution},
  author={Mao, Yuping and Chaudhary, Manish and Kondappan, Manikandan and Shi, Junheng and Ilo-Okeke, Ebubechukwu O and Ivannikov, Valentin and Byrnes, Tim},
  journal={Phys. Rev. Lett.},
  volume={131},
  number={11},
  pages={110602},
  year={2023},
  publisher={APS}
}

@article{kirby2023exact,
  title={Exact and efficient {Lanczos} method on a quantum computer},
  author={Kirby, William and Motta, Mario and Mezzacapo, Antonio},
  journal={Quantum},
  volume={7},
  pages={1018},
  year={2023},
  publisher={Verein zur F{\"o}rderung des Open Access Publizierens in den Quantenwissenschaften}
}

@article{kingma2014adam,
  title={Adam: A method for stochastic optimization},
  author={Kingma, Diederik P and Ba, Jimmy},
  eprint={1412.6980},
  archivePrefix={arXiv},
  year={2014},
  note={arXiv preprint 1412.6980},
  doi={10.48550/arXiv.1412.6980},
  primaryClass={cs.LG}
}

@article{byrd1995limited,
  title={A limited memory algorithm for bound constrained optimization},
  author={Byrd, Richard H and Lu, Peihuang and Nocedal, Jorge and Zhu, Ciyou},
  journal={SIAM J. Sci. Comput.},
  volume={16},
  number={5},
  pages={1190--1208},
  year={1995},
  publisher={SIAM},
  doi={10.1137/0916069},
}

@article{kuhn2023bayesian,
  title={Bayesian parameterization of continuum battery models from featurized electrochemical measurements considering noise},
  author={Kuhn, Yannick and Wolf, Hannes and Latz, Arnulf and Horstmann, Birger},
  journal={Batter. Supercaps},
  volume={6},
  number={1},
  pages={e202200374},
  year={2023},
  publisher={Wiley Online Library}
}

@article{costa2025further,
  title={Further improving quantum algorithms for nonlinear differential equations via higher-order methods and rescaling},
  author={Costa, Pedro CS and Schleich, Philipp and Morales, Mauro ES and Berry, Dominic W},
  journal={npj Quantum Inf.},
  volume={11},
  number={1},
  pages={141},
  year={2025},
  publisher={Nature Publishing Group UK London}
}

@article{sarma2024quantum,
  title={Quantum variational solving of nonlinear and multidimensional partial differential equations},
  author={Sarma, Abhijat and Watts, Thomas W and Moosa, Mudassir and Liu, Yilian and McMahon, Peter L},
  journal={Phys. Rev. A},
  volume={109},
  number={6},
  pages={062616},
  year={2024},
  publisher={APS}
}

@article{berger2025trainable,
  title={Trainable embedding quantum physics informed neural networks for solving nonlinear {PDEs}},
  author={Berger, Stefan and Hosters, Norbert and M{\"o}ller, Matthias},
  journal={Sci. Rep.},
  volume={15},
  number={1},
  pages={18823},
  year={2025},
  publisher={Nature Publishing Group UK London}
}

@article{tennie2025integration,
  title={Integration of the {Fokker--Planck} equation on quantum computers: A new path to modelling nonlinear dynamics},
  author={Tennie, Felix and Magri, Luca},
  journal={Proc. R. Soc. A},
  volume={481},
  number={2326},
  pages={20250016},
  year={2025},
  publisher={The Royal Society}
}

@article{jin2024analog,
  title={Analog quantum simulation of partial differential equations},
  author={Jin, Shi and Liu, Nana},
  journal={Quantum Sci. Technol.},
  volume={9},
  number={3},
  pages={035047},
  year={2024},
  publisher={IOP Publishing}
}

@article{feulner2026quantum,
  title={Quantum gates with parametrically driven multi-qubit couplers},
  author={Feulner, Verena and Fani, Marjan and Heunisch, Lukas and Tasler, Stephan and Hartmann, Michael J},
  note={arXiv preprint arXiv:2606.14522},
  doi={10.48550/arXiv.2606.14522},
  year={2026}
}

@article{renger2026superconducting,
  title={Superconducting qubit-resonator quantum processor with effective all-to-all connectivity},
  author={Renger, Michael and Verjauw, Jeroen and Wurz, Nicola and Hosseinkhani, Amin and Ockeloen-Korppi, Caspar and Liu, Wei and Rath, Aniket and Thapa, Manish J and Vigneau, Florian and Wybo, Elisabeth and others},
  journal={Phys. Rev. Res.},
  volume={8},
  number={1},
  pages={013148},
  year={2026},
  publisher={APS}
}

@article{campbell2017roads,
  title={Roads towards fault-tolerant universal quantum computation},
  author={Campbell, Earl T and Terhal, Barbara M and Vuillot, Christophe},
  journal={Nature},
  volume={549},
  number={7671},
  pages={172--179},
  year={2017},
  publisher={Nature Publishing Group UK London}
}

@article{abdurakhimov2024technology,
  title={Technology and performance benchmarks of IQM's 20-qubit quantum computer},
  author={Abdurakhimov, Leonid and Adam, Janos and Ahmad, Hasnain and Ahonen, Olli and Algaba, Manuel and Alonso, Guillermo and Bergholm, Ville and Beriwal, Rohit and Beuerle, Matthias and Bockstiegel, Clinton and others},
  note={arXiv preprint arXiv:2408.12433},
  doi={10.48550/arXiv.2408.12433},
  year={2024}
}

@article{braun2015thermodynamically,
author = {Braun, Stefanie and Yada, Chihiro and Latz, Arnulf},
title = {Thermodynamically consistent model for space-charge-layer formation in a solid electrolyte},
journal = {J. Phys. Chem. C},
volume = {119},
number = {39},
pages = {22281-22288},
year = {2015},
doi = {10.1021/acs.jpcc.5b02679},
URL = { 
        https://doi.org/10.1021/acs.jpcc.5b02679
},
eprint = { 
        https://doi.org/10.1021/acs.jpcc.5b02679
}
}

@article{thomas2016classical,
  title={Classical magnetic dipole moments for the simulation of vibrational circular dichroism by ab initio molecular dynamics},
  author={Thomas, Martin and Kirchner, Barbara},
  journal={J. Phys. Chem. Lett.},
  volume={7},
  number={3},
  pages={509--513},
  year={2016},
  publisher={ACS Publications}
}

@article{hein2016stochastic,
  title={Stochastic microstructure modeling and electrochemical simulation of lithium-ion cell anodes in 3D},
  author={Hein, Simon and Feinauer, Julian and Westhoff, Daniel and Manke, Ingo and Schmidt, Volker and Latz, Arnulf},
  journal={J. Power Sources},
  volume={336},
  pages={161--171},
  year={2016},
  publisher={Elsevier}
}

@article{javadi2024quantum,
  title={Quantum computing with Qiskit},
  author={Javadi-Abhari, Ali and Treinish, Matthew and Krsulich, Kevin and Wood, Christopher J and Lishman, Jake and Gacon, Julien and Martiel, Simon and Nation, Paul D and Bishop, Lev S and Cross, Andrew W and others},
  note={arXiv preprint arXiv:2405.08810},
  doi={10.48550/arXiv.2405.08810},
  year={2024}
}

@article{nation_scalable_2021,
	title = {Scalable mitigation of measurement errors on quantum computers},
	volume = {2},
	issn = {2691-3399},
	url = {https://link.aps.org/doi/10.1103/PRXQuantum.2.040326},
	doi = {10.1103/PRXQuantum.2.040326},
	number = {4},
	urldate = {2022-10-11},
	journal = {PRX Quantum},
	author = {Nation, Paul D. and Kang, Hwajung and Sundaresan, Neereja and Gambetta, Jay M.},
	month = nov,
	year = {2021},
	pages = {040326},
}

@article{wu2026integrative,
  title={Integrative additive design for robust {SEI} formation in {NMC811}||silicon batteries},
  author={Wu, Xianyang and Li, Xinlin and Yang, Zhenzhen and Ingram, Brian J and Li, Matthew and Su, Chi-Cheung and Amine, Khalil},
  journal={ACS Appl. Mater. Interfaces},
  volume={18},
  number={2},
  pages={3714--3722},
  year={2026},
  publisher={ACS Publications}
}

@article{gubaev2026variational,
  title={Variational quantum algorithm for anion exchange across an electrolyzer membrane},
  author={Gubaev, Timur and Pfeffer, Philipp and Dre{\ss}ler, Christian and Schumacher, J{\"o}rg},
  journal={Phys. Rev. Appl.},
  volume={26},
  number={1},
  pages={014051},
  year={2026},
  publisher={APS}
}

@article{lehnert2025combining,
  title={Combining Molecular Dynamics and Experimental Methods for the Parametrization of Binary Carbonate-Based Electrolytes},
  author={Lehnert, Lukas and Lorenz, Martin and Juarez, Maria Fernanda and Schammer, Max and Nojabaee, Maryam and Sch{\"o}nhoff, Monika and Horstmann, Birger},
  journal={Journal of The Electrochemical Society},
  volume={172},
  number={5},
  pages={050523},
  year={2025},
  publisher={IOP Publishing}
}

\end{document}